\documentclass[sigconf,9pt]{acmart}
\usepackage{algorithm}
\usepackage{algpseudocode}
\usepackage{enumitem}
\AtBeginDocument{%
  }

\AtBeginEnvironment{align}{%
  \setlength{\abovedisplayskip}{1.5pt}%
  \setlength{\belowdisplayskip}{1.5pt}
}
\makeatletter
\AtBeginDocument{%
  \fancypagestyle{standardpagestyle}{%
    \fancyhf{}%
  }%
  \fancypagestyle{firstpagestyle}{%
    \fancyhf{}%
  }%
  \pagestyle{empty}%
}
\makeatother

\copyrightyear{2026}
\acmYear{2026}
\setcopyright{cc}
\setcctype{by}
\acmConference[DAC '26]{63rd ACM/IEEE Design Automation Conference}{July 26--29, 2026}{Long Beach, CA, USA}
\acmBooktitle{63rd ACM/IEEE Design Automation Conference (DAC '26), July 26--29, 2026, Long Beach, CA, USA}
\acmDOI{10.1145/3770743.3804196}
\acmISBN{979-8-4007-2254-7/2026/07}

\begin{document}

\title{Compiler Framework for Directional Transport in Zoned Neutral Atom Systems with AOD Assistance: A Hybrid Remote CZ Approach}

% \affiliation{%

% \affiliation{%

% \affiliation{%

% \affiliation{%

% \affiliation{%

% \affiliation{%

\author{}
\makeatletter
\gdef\authors{Lingyi Kong, Chen Huang, Zhemin Zhang, Yidong Zhou, Xiangyu Ren, Shaochen Li, and Zhiding Liang}
\makeatother
\renewcommand{\shortauthors}{Kong et al.}

\makeatletter
\def\@mkauthors{%
  \global\setbox\mktitle@bx=\vbox{%
    \noindent\unvbox\mktitle@bx\par
    \vskip 0.5em
    \centering
    {\Large
      Lingyi Kong$^{\S 1}$, Chen Huang$^{\S 1}$, Zhemin Zhang$^{\S}$, 
      Yidong Zhou$^{\dagger}$, Xiangyu Ren$^{\ddagger}$, 
      Shaochen Li$^{\S}$, Zhiding Liang$^{\S *}$\par
    }
    \vskip 0.35em
    {\normalsize
      $^{\S}$The Chinese University of Hong Kong \quad
      $^{\dagger}$Rutgers University \quad
      $^{\ddagger}$The University of Edinburgh\par
    }
    {\normalsize
      $^{*}$Corresponding author: \texttt{zliang@cse.cuhk.edu.hk}\par
    }
    {\normalsize
      $^{1}$These authors contributed equally to this work.\par
    }
    \vskip 0.6em
  }%
}
\def\@authorsaddresses{}
\def\@authornotes{}
\makeatother

\begin{abstract}
% Zoned neutral-atom arrays separate storage, entanglement, and readout to boost fidelity. However, long-range entanglement remains movement-bound: AOD shuttling faces row/column non-crossing, device-speed limits, and an FOV/NA-limited range. We present a compiler for zoned architectures that makes directional transport the default for remote \(\mathrm{CZ}\), reserving AODs for channel setup and micro-tuning. Under antiblockade, a detuning-modulated  \(\pi\)-pulse sequence drives directional transport of a Rydberg excitation along a dynamic and resettable ancilla corridor, realizing a \(\mathrm{CZ}\) gate between stationary, non-adjacent qubits. Our approach cuts entangling-stage duration by approximately \(50\%\) versus AOD-only baselines and enables long-distance connectivity beyond objective-limited shuttling.

We present a directional-transport (DT)-based remote CZ gate and compiler for zoned neutral-atom arrays that overcomes movement-bound entanglement limitations. Current AOD-based shuttling faces row/column non-crossing constraints, device-speed limits, and hardware-restricted range—bottlenecks for long-distance connectivity. Our approach reserves AODs for channel setup and micro-tuning while making DT the default for remote entanglement. Under antiblockade, a detuning-modulated $\pi$-pulse sequence drives directional transport of a Rydberg excitation along a dynamic and resettable ancilla corridor, realizing a CZ gate between stationary, non-adjacent qubits. This cuts entangling-stage duration by approximately 50\% -90\% versus AOD-only baselines and enables long-distance connectivity beyond objective-limited shuttling.

\end{abstract}

\maketitle

\section{Introduction}
Neutral atom (NA) arrays have emerged as a leading platform for building scalable quantum processors by trapping individual atoms in optical tweezers and leveraging Rydberg interactions~\cite{manetsch2025tweezer, graham2022multi, schmid2024hybrid, nottingham2023decomposing,stade2024abstract,PRXQuantum1,optica1,science_na,bluvstein2024logical}. However, realizing practical quantum computation at scale requires efficient compilation strategies that minimize the substantial overhead of atomic transport.

In zoned neutral atom architectures, long-range entanglement is typically achieved by physically shuttling qubits between storage and entanglement zones using acousto-optic deflectors (AODs), executing local two-qubit gates, and shuttling them back~\cite{lin2025reuse, huang2024zap, Geyser,decker2024arcticfieldprogrammablequantum,Beugnon,nature3,Urban_2009,science_mq,iccad_qmra,tan2025compilation,tan2024compiling, Atomique,10082942,brandhofer2021optimalmappingneartermquantum, CGO25,powermove, Schmid_2024}. While Rydberg-based gates themselves operate on nanosecond to microsecond timescales, mechanical motion between zones routinely consumes tens to hundreds of microseconds per round trip. For circuits with dense long-range entanglement patterns such as the quantum Fourier transform, this transport overhead becomes dominant, making AOD motion rather than gate execution the primary latency bottleneck.

\begin{figure}[ht]
    \centering
    \includegraphics[width=0.97\linewidth]{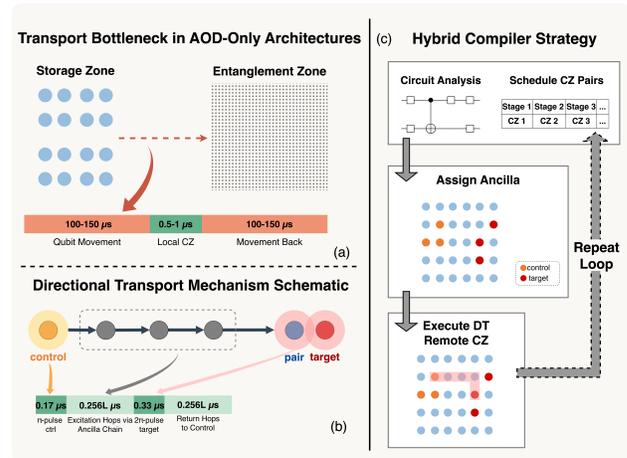}
    \caption{Directional-transport compilation framework.}

    \label{fig:teaser}
    \vspace{-2mm}
\end{figure}

Directional transport (DT) offers a fundamentally different approach by propagating quantum information as a coherent Rydberg excitation along a pre-configured chain of ancilla atoms, rather than by moving physical atoms~\cite{isenhower2010demonstration, wang2025directional, valencia2024rydberg}. The mechanism exploits Rydberg antiblockade, wherein a precisely tuned sequence of laser pulses swaps population between adjacent atoms through facilitation. A remote CZ gate proceeds as follows: a local $\pi$-pulse maps the $|1\rangle_c$ component of the control qubit into a single Rydberg excitation, this excitation is transported via DT to a relay atom adjacent to the target, and a $2\pi$-pulse on the target tuned to the antiblockade resonance imprints a conditional phase only when the neighboring excitation is present. As shown in Fig.~\ref{fig:teaser}(a)(b), with realistic parameters, such remote CZ gates complete on timescales of a few microseconds, compared to tens or hundreds of microseconds for AOD-only approaches. This one to two order-of-magnitude speedup motivates using directional transport, provided compilation overhead can be minimized.

The central challenge is that maintaining static DT corridors to cover all possible interactions would consume prohibitive physical space and AOD overhead, negating the speedup benefits~\cite{baker2021exploiting}. Effective compilation must therefore integrate AOD transport with dynamic DT channel planning, using AOD motion strategically for initial configuration and maintenance rather than for every entangling operation~\cite{wang2024q}.

We propose a DT-aware compiler framework that achieves this balance 
through intelligent CZ pair classification and dynamic channel management; 
the framework is illustrated in Fig.~\ref{fig:teaser}(c). The framework performs AOD-assisted rearrangement in an initial configuration phase, concentrating bulk atomic motion into this stage. Qubits are assigned to pre-defined DT channels using priority-based placement, and a conflict-graph router with greedy maximal independent set scheduling parallelizes AOD movements. Once the topology is established, the circuit executes on top of an essentially static infrastructure. Every CZ pair is classified as either DT-eligible or direct-AOD: pairs that align cheaply to one-dimensional DT channels proceed via fast remote CZ, while remaining pairs use local AOD shuttling. By incrementally maintaining a cost-aware DT backbone that grows and reshapes across stages rather than repeatedly repacking all data qubits, and by reusing a small pool of flying ancilla as relay sites, the framework replaces large global AOD sweeps with localized updates. This approach substantially reduces AOD overhead while enabling complex long-range entanglement patterns.

The contributions of this work are threefold: 

\begin{itemize}[topsep=0pt, partopsep=0pt, itemsep=2pt, parsep=0pt, leftmargin=*]
    \item We introduce a hybrid compilation methodology that tightly couples AOD transport with dynamic DT channel planning, showing how to exploit directional transport benefits while keeping resource overhead manageable for practical circuits.
    \item We develop specialized algorithms for QFT-style circuits with low parallelism and sliding interaction patterns, where dynamic channel management provides advantages over static global provisioning.
    \item We demonstrate that our approach achieves approximately 50\% -90\% reduction in entangling-stage duration compared to AOD-only baselines while enabling long-distance connectivity beyond the reach of field-of-view and numerical aperture limitations of optical systems.
\end{itemize}

\section{Background}

% As classical computing systems face fundamental scaling challenges for certain problem classes, quantum computing has emerged as a powerful alternative. By leveraging the quantum-mechanical principles of superposition and entanglement, quantum computers can, in principle, explore vast computational state spaces to solve problems intractable for even the largest classical supercomputers.

% A quantum bit, or ``qubit,'' is the fundamental unit of quantum information. Unlike a classical bit, which is restricted to a state of either 0 or 1, a qubit can exist in a coherent superposition of both states simultaneously. The true power of quantum computation, however, arises from entanglement: a unique correlation between multiple qubits that allows the system's state to be described only as a whole, not by its independent parts.

% Many physical platforms are being explored to build quantum processors; NA arrays have become a leading candidate. In this modality, individual atoms are trapped in a vacuum by focused laser beams called optical tweezers. These arrays offer compelling advantages, particularly in scalability to large numbers of high-quality qubits and the ability to engineer strong, switchable, long-range interactions by exciting atoms to highly energetic Rydberg states.

\subsection{AOD-based Shuttling and Traditional Zoned NA Compilers}
A central design pattern for scaling NA systems is the zoned architecture, in which the QPU is partitioned into storage zones for long-coherence idling, entanglement zones for gate execution, and readout zones for measurement. Moving qubits across these regions relies on AOD-based shuttling, where optical tweezers are steered to transport atoms within the array. Remote entangling operations typically require bringing two qubits from storage into the entanglement zone, applying a local gate, and returning them afterward. This physical motion is the dominant performance bottleneck: AOD shuttling takes tens to hundreds of microseconds, far slower than the nanosecond-to-microsecond timescale of Rydberg gates, and is constrained by limited field-of-view and routing restrictions such as non-crossing tweezer trajectories.

\subsection{Directional Transport}

AOD shuttling moves the physical atom, an alternative mechanism for conveying quantum information is DT, which moves only a quantum state, specifically, a single Rydberg excitation.

This is achieved using a pre-arranged, staitc channel of ancilla atoms. The transport relies on a phenomenon known as Rydberg antiblockade or facilitation. An atom in a Rydberg state shifts the energy levels of its neighbors. By applying a precisely tuned sequence of laser pulses (e.g., $\pi$-pulses), an excitation on one atom can be coherently and directionally swapped to its neighbor, but only if its neighbor is already excited. This creates a ``domino effect,'' where a single excitation can be rapidly ``hopped'' along the ancilllary chain from a control qubit to a target qubit.

DT is fast, each hop is driven by fast laser pulses and can be completed on the sub-microsecond timescale, bypassing the slow mechanical limitations of AODs entirely. However, DT requires a static, pre-configured ``corridor'' of ancilla atoms. Thus, our work exploits this trade-off, proposing a hybrid compiler framework that uses AODs once to configure these fast DT channels, and then leverages DT for all subsequent remote entanglement. 

% \section{DT-based Remote CZ} \zd{}
\section{Motivation}

\begin{figure}
  \centering
  \includegraphics[width=0.94\linewidth]{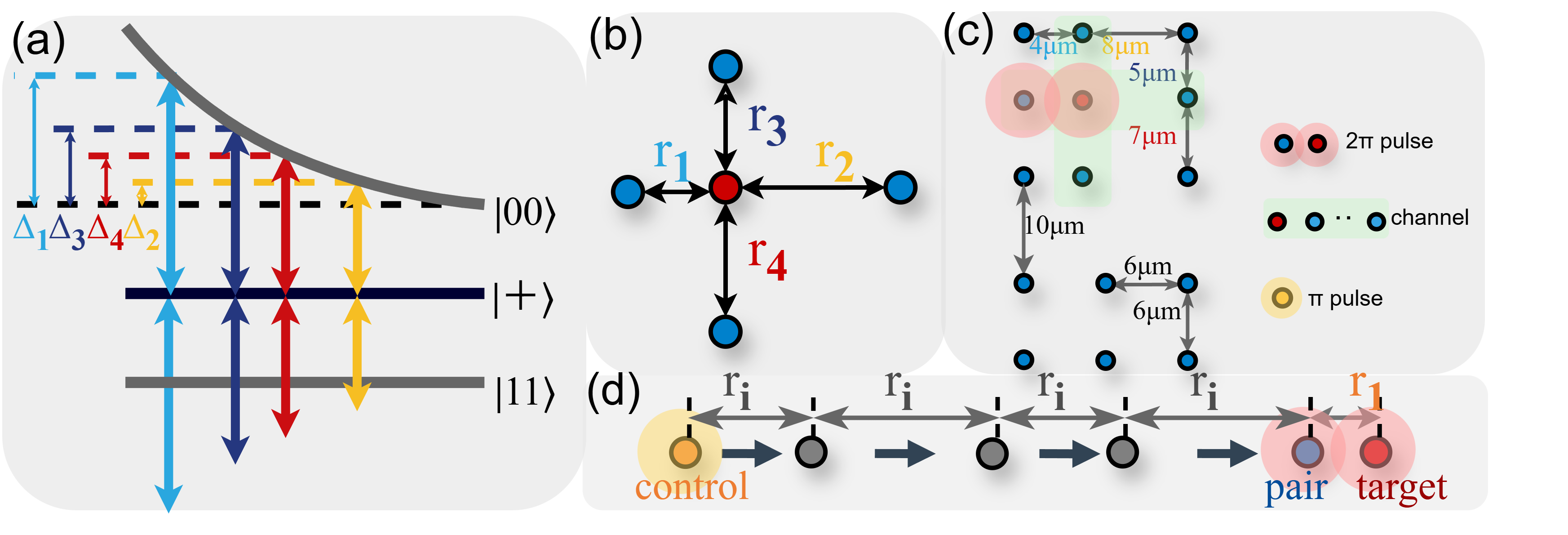}
  \caption{Remote CZ gate and compiler architecture. 
(a) Energy-level diagram of the conditional $2\pi$ antiblockade pulse on the target: the interaction potential (blue curve) shifts the levels so that the $2\pi$ pulse is resonant only at the facilitation distance. 
(b) Extension of the 1D DT mechanism to a 2D geometry. 
(c) Representative geometric constraints used by the compiler, including entanglement-zone channel spacing (4, 8, 5, 7~$\mu$m), zone separation (10~$\mu$m), and storage-zone spacing (6~$\mu$m). 
(d) Temporal sequence of the remote CZ protocol: a $\pi$ pulse (yellow) on the control creates a Rydberg excitation that propagates along the ancilla chain, enabling a conditional $2\pi$ pulse on the target(red).}
  \Description{}
  \label{fig:theory}
  \vspace{-2.ex}
\end{figure}

In zoned neutral-atom processors, the latency of long-range entanglement is dominated not by the Rydberg CZ itself but by the slow physical motion required to move atoms between storage and entangling zones. Although a local Rydberg CZ can be executed in sub-microsecond time, AOD-based shuttling typically incurs tens to hundreds of microseconds per round trip. As in SWAP-based remote CZ in gate-model circuits, the routing rather than the entangling pulse, quickly becomes the primary bottleneck. DT mitigates this imbalance by eliminating most of the expensive motion. Instead of rearranging atoms or inserting long SWAP ladders, DT keeps data qubits stationary and propagates a single Rydberg excitation through a facilitation channel via microsecond-scale hops. A DT-mediated remote CZ routinely completes within $1$--$10\,\mu\mathrm{s}$, providing a one- to two-order-of-magnitude speedup over AOD-only implementations and significantly reducing idle-error exposure for large-scale and early fault-tolerant workloads.

DT does not replace AOD; the two mechanisms are tightly coupled. All Rydberg operations still occur inside the entanglement zone, and DT channels must be pre-constructed by transporting ancilla and selected qubits onto channel sites. AOD performs zone-to-zone movement and channel construction, while DT executes the fast excitation transfer along the established path. The DT-aware static compiler introduced next is designed to coordinate this joint AOD+DT workflow, concentrating slow transport into an early configuration stage so that subsequent circuit execution is dominated by fast DT hops and local Rydberg pulses.
\begin{figure*}[!t]
    \centering
    \includegraphics[width=0.97\textwidth]{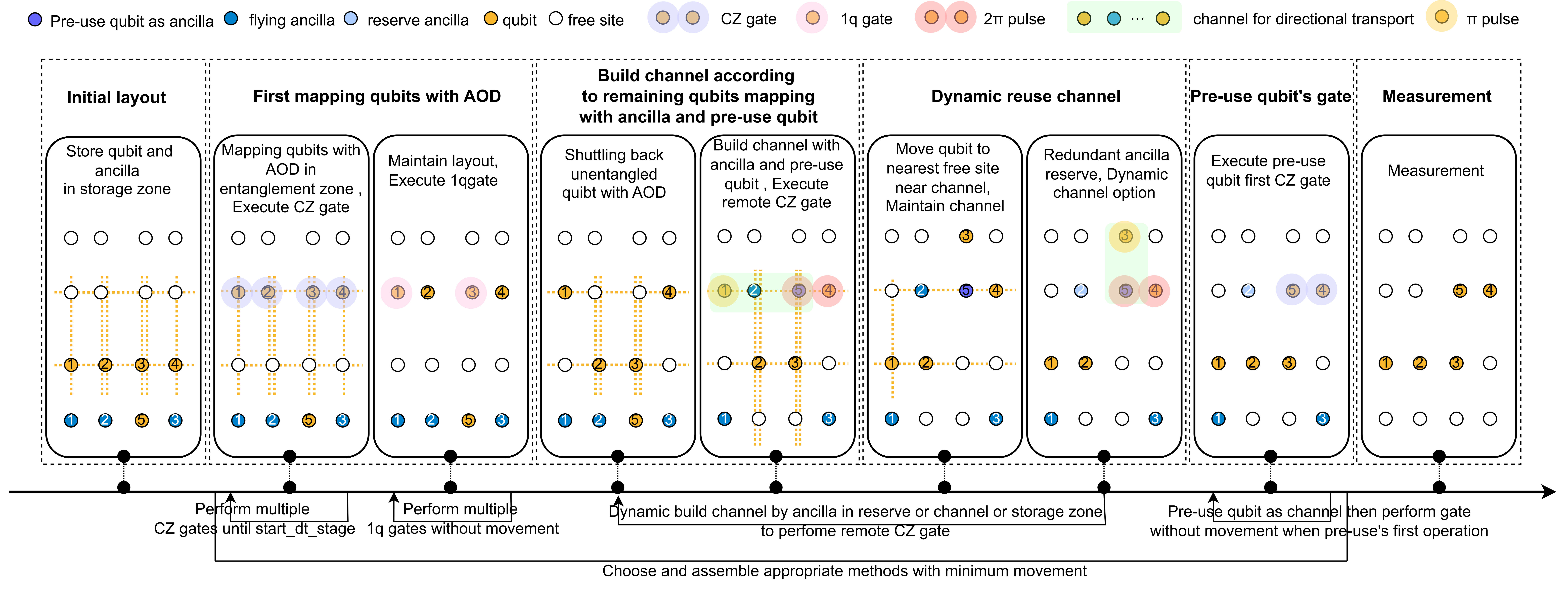}
   \caption{Flowchart of the customized dynamic compiler framework. }
    \label{fig:fpqa-flow}
    \vspace{-2.ex}
\end{figure*}

\section{Compiler Framework for Hybrid Remote CZ}
\label{sec:framework}
In this section, we turn DT–based remote CZ into a concrete hybrid compilation framework for zoned architectures as shown in Fig. \ref{fig:fpqa-flow}.
\subsection{General Theory of Remote CZ by Directional Transport}

% Building on the concept of DT~\cite{wang2025directional}, we propose the general theory of achieving the remote CZ by DT, which realizes an entangling gate between a control qubit \(c\) and a target qubit \(t\) by propagating a \emph{single Rydberg excitation} rather than the logical qubit states themselves, along a chain of ancilla atoms. 
Building on the concept of DT~\cite{wang2025directional}, we propose the general theory of achieving the remote CZ by DT as illustrated in Fig.~\ref{fig:theory}. Instead of physically moving the logical qubits, the entangling interaction is mediated by propagating a \emph{single Rydberg excitation} along a chain of ancilla atoms from the control qubit \(c\) and the target qubit \(t\).
Throughout this work, \(\lvert 0\rangle_c\) and \(\lvert 1\rangle_c\) denote hyperfine ground states of the control qubit, and \(\lvert r\rangle_c\) denotes a highly excited Rydberg state used for entangling operations. The protocol begins with a local ``map-in'' \(\pi\)-pulse resonant with the \(\lvert 1\rangle_c \leftrightarrow \lvert r\rangle_c\) transition: only the computational state \(\lvert 1\rangle_c\), encoded in a ground hyperfine level, is driven to the Rydberg excited state \(\lvert r\rangle_c\), whereas \(\lvert 0\rangle_c\) remains in the ground-state manifold. In this way, the logical information is encoded in the amplitudes \(\alpha,\beta\) of the control qubit, while the \emph{presence or absence} of a Rydberg excitation serves as a binary flag for the subsequent transport. Denoting the Rabi frequency on this transition by \(\Omega\), we take a representative value \(\Omega/(\pi)\approx 3~\mathrm{MHz}\), which corresponds to a \(\pi\)-pulse duration \(t_{\pi}\approx 0.167~\mu\mathrm{s}\).

% The single Rydberg excitation is then relayed along a DT chain of length \(L\) by repeating a nearest-neighbor ``hop'' operation. Each hop consists of a short detuning switch followed by a pair of facilitated \(\pi\)-pulses that swap population \(\lvert r,0\rangle \leftrightarrow \lvert 0,r\rangle\) between adjacent ancilla atoms, while all other sites remain in their ground states. 
% The excitation is then relayed along a DT chain of length \(L\) by repeating a nearest-neighbor ``hop'' operation. Each hop consists of a short detuning switch followed by a pair of facilitated \(\pi\)-pulses that swap population \(\lvert r,0\rangle \leftrightarrow \lvert 0,r\rangle\) between adjacent ancilla atoms, while all other sites remain in their ground states. 
% Denoting the effective Rabi frequency on the bright superposition by \(\tilde{\Omega}\) and the detuning-switch time by \(\tau_{\mathrm{sw}}\), the corresponding hop duration is \(T_{\mathrm{hop}} = 2\pi/\tilde{\Omega} + \tau_{\mathrm{sw}}\), which in our working parameter regime is approximately \(T_{\mathrm{hop}} \approx 0.256~\mu\mathrm{s}\). After \(L\) such hops, the Rydberg excitation resides on a relay atom adjacent to the target qubit \(t\), while both \(c\) and \(t\) themselves remain in the ground-state manifold.
The single Rydberg excitation is then relayed along a DT chain of length \(L\) by repeating a nearest-neighbor hop operation. Each hop consists of a short detuning switch followed by a pair of facilitated \(\pi\)-pulses that swap population \(\lvert r,0\rangle \leftrightarrow \lvert 0,r\rangle\) between adjacent ancilla atoms, while all other sites remain in their ground states. Denoting the effective Rabi frequency on the bright superposition by \(\tilde{\Omega}\) and the detuning switch time by \(\tau_{\mathrm{sw}}\), the hop duration is \(T_{\mathrm{hop}} = 2\pi/\tilde{\Omega} + \tau_{\mathrm{sw}}\), which in our working parameter regime is approximately \(T_{\mathrm{hop}} \approx 0.256~\mu\mathrm{s}\)~\cite{wang2025directional}. After \(L\) such hops, the Rydberg excitation resides on a relay atom adjacent to the target qubit \(t\), while both \(c\) and \(t\) remain in the ground state manifold.

At this point, we apply a single Rydberg \(2\pi\)-pulse on the \(\lvert 1\rangle_t \leftrightarrow \lvert r\rangle_t\) transition of the target, where \(\lvert 0\rangle_t\) and \(\lvert 1\rangle_t\) are again hyperfine ground states and \(\lvert r\rangle_t\) is a Rydberg excited state. When the neighboring relay atom is in \(\lvert r\rangle\), the Rydberg--Rydberg interaction shifts the target transition into an antiblockade (facilitation) resonance. The resulting \(2\pi\) evolution imprints an additional conditional phase on the \(\lvert 1\rangle_t\) component, thereby realizing a CZ gate on \((c,t)\) up to single-qubit phases. The duration of this entangling pulse is set by the same Rabi frequency \(\Omega\); for a representative value \(\Omega/(2\pi)\approx 3~\mathrm{MHz}\), the corresponding \(2\pi\)-pulse time is \(t_{2\pi}\approx 0.334~\mu\mathrm{s}\), which is of the same order as (and roughly twice) \(t_{\pi}\).

After the entangling step, the Rydberg excitation is transported back to the control side by reversing the DT sequence, incurring another time cost of \(L T_{\mathrm{hop}}\). A second local \(\pi\)-pulse on \(\lvert r\rangle_c \rightarrow \lvert 1\rangle_c\) then maps the excitation from the Rydberg manifold back into the ground-state computational level, returning the control qubit entirely to the ground manifold. Residual dynamical phases on both \(c\) and \(t\) are absorbed into virtual \(Z\) rotations implemented in software, with negligible temporal overhead. Collecting all contributions, the total duration of a DT-mediated remote CZ gate is
\begin{align}
T_{\mathrm{CZ}} \;\simeq\; 2 t_{\pi} + 2 L T_{\mathrm{hop}} + t_{2\pi}
\;\approx\; 0.667~\mu\mathrm{s} + 0.512\,L~\mu\mathrm{s}.
\label{eq:duration}
\end{align}

Thus, the cost of remote entanglement scales linearly with the number of fast excitation hops rather than with slow mechanical motion, aligning the physical implementation of remote CZ with the fast timescales of native Rydberg gates.

\subsection{Compiler Support for DT-based Remote CZ}
All logical CZ gates are realized through DT-based remote CZ, eliminating most repeated shuttling between storage and entanglement zones. The compiler performs a single AOD-assisted configuration step to place qubits onto a set of predefined DT channels inside the entanglement zone. Once this topology is established, entanglement is achieved by routing DT excitations along channels, overall execution time could reduced.

Our approach shares several algorithmic elements with prior zoned compilers such as ZAC~\cite{lin2025reuse} and ZAP~\cite{huang2024zap}, including explicit modeling of AOD motion, use of movement vectors, and MIS-based parallel scheduling of rearrangements. Like ZAP, we support large-scale AOD-assisted reconfiguration for FPQA-style devices. However, unlike ZAC and ZAP, which repeatedly repartition qubits and schedule multiple rounds of zone-to-zone shuttling across successive Rydberg ``stages’’, our compiler relies on an essentially static DT topology and executes all one- and two-qubit gates in the temporal order produced by the logical scheduler. DT-based remote CZ handles all non-local interactions, so any further motion is confined to small, local refinements within the entanglement zone, with the bulk of the transport cost absorbed upfront.

Each DT channel is realized as a one-dimensional chain of \emph{flying ancilla} atoms. This concept follows FPQA compilers such as Q-Pilot~\cite{wang2024q}, where movable AOD-trapped atoms act as temporary routing resources that can be created, positioned, and recycled on demand. In our setting, flying ancilla similarly serve as transport carriers for DT propagation, enabling non-local CZ operations without relocating the data qubits themselves.

\noindent\textbf{Priority-Based Placement and Static DT.}
We first assign a priority score $\text{Pri}(q)$ to each qubit. To this end, we traverse the logically scheduled circuit in temporal order, and for every gate appearing at a given time layer $p$, add a layer weight $\mathrm{Weight}_p = 1/(p+1)$ to all participating qubits. Sorting qubits in descending order of the accumulated weight yields a ranking in which qubits that appear earlier and more frequently in the computation are given higher priority. The hardware SLM grid is then adaptively cropped. In the storage region, we estimate the required number of columns and rows as a function of the number of qubits and gradually increase the grid size until it can accommodate all qubits. In the entanglement region, the column/row range is derived from the storage-region size in order to support the desired parallelism and pairing structure. To respect hardware layout constraints, we restrict qubit sites in the entanglement zone to a fixed subset of columns (e.g., those with $x \bmod 12 = 4$), and denote the first usable entangling row in these columns by $ent\_first\_row\_y$. As illustrated in Fig.~\ref{fig:theory}(c), these hardware constraints define the physical layout, such as typical entanglement-zone atom separations (e.g., 4–8 $\mu$m), storage-zone spacing (e.g., 6 $\mu$m), and inter-zone gaps (e.g., 10 $\mu$m).

From candidate sites at and below this row, we then select as many positions as there are qubits (with appropriate column spacing) as target locations in the entanglement zone and sort them by coordinate. Target sites are assigned in a priority-first manner: following the order of $\text{Pri}(q)$, each qubit is greedily matched to the next available target site, ensuring that critical qubits receive better entanglement-zone coordinates. To reduce subsequent migration overhead, the initial storage locations of qubits are chosen by minimizing a column-dominant placement cost whose main term penalizes column mismatch and whose secondary terms encourage row alignment and proximity to the first entangling row. This cost structure significantly reduces both the average storage-to-entanglement distance and the likelihood of routing conflicts and effectively aligns important qubits with the heads of their associated DT channels. In contrast to ZAC’s reuse-aware placement, which is applied around multiple entangling windows to exploit qubit reuse, our placement is computed once before execution and is explicitly tailored to a static DT topology that will be reused throughout the entire circuit.

\noindent \textbf{DT-Channel Configuration and Bulk Migration.}
To enable DT-based CZ, each qubit column is aligned with a static DT channel that supplies the facilitation sites required for $2\pi$ pulses and for forming connected DT paths across qubits. Given the qubit endpoints in the entanglement zone, the compiler identifies for each column its first-row entanglement site and paired column, merging them into a set of non-overlapping DT-channel targets. A symmetric pairing map is precomputed to guarantee consistent connectivity and ensure that every qubit can reach any other qubit through relay atoms along at least one DT path. On the storage side, SLM sites already occupied by qubits are excluded, and the same column-oriented cost model used during placement is applied to choose channel starting locations. This effectively pre-allocates DT channels in the entanglement zone and prepares the corresponding storage-to-entanglement move vectors, allowing channels to be populated during an early configuration phase without interfering with qubit motion. Unlike ZAP’s simulated-annealing–based intermediate placements that are recomputed for different entangling segments, our DT-channel configuration is computed once and treated as a static routing resource throughout execution.

% To support DT-based CZ, each qubit column is positioned adjacent to a static directional-transport (DT) channel that provides the physical sites required for facilitation-based $2\pi$ pulses and for realizing connected DT paths between qubits. Starting from the previously determined qubit targets in the entanglement zone, the compiler derives, for every qubit column, the first-row entanglement site and its paired column, and merges these into a set of DT-channel target sites that are guaranteed not to overlap with qubit endpoints. A symmetric pairing map in the entanglement region is precomputed to ensure consistent connectivity along these channels and to guarantee that every qubit has access to at least one DT path to any other qubit via relay atoms.

% On the storage side, SLM sites already occupied by qubits are removed from consideration, and the same column-dominant cost function used for placement is applied to select starting locations for DT-channel resources. Conceptually, this step pre-allocates DT channels in the entanglement zone and prepares the corresponding storage-to-entanglement move vectors, so that channels can be populated in an early configuration phase before two-qubit interactions are performed, without introducing additional conflicts with qubits. In contrast to ZAP’s simulated-annealing-based intermediate placements that are recomputed for different entangling segments of the circuit, our DT-channel configuration is computed once and then treated as a static resource that the scheduler can rely on throughout execution.

\noindent\textbf{Parallel Routing and Duration Modeling.}
Given the initial placement and DT-channel layout, the compiler performs a bulk-oriented migration phase before any two-qubit operations. The router translates the delta between the current and target mappings into hardware-level movement primitives (activate, move/park, deactivate). We build a conflict graph over both qubit and DT-channel moves, where vertices represent move vectors and edges capture collisions or AOD-channel contention. At each routing step, a greedy maximal-independent-set (MIS) pass selects a large set of conflict-free moves that can run in parallel, maximizing AOD utilization. When a single-pass resolution is impossible, the router falls back to a lightweight multi-stage sequence (Activate $\to$ Park $\to$ BigMove $\to$ Park $\to$ Deactivate), temporarily parking qubits on auxiliary SLM rows. This staged procedure guarantees collision-free motion and avoids hardware-level deadlock.
After the initial configuration, the compiler keeps the qubit mapping largely static. Instead of repeatedly repartitioning the shuttling qubits between storage and entanglement zones for interaction windows, we follow the logical schedule directly on a mostly fixed DT topology, using remote CZ operations for all non-local gates. Any remaining motion is restricted to small, circuit-specific adjustments inside the entanglement zone, with the bulk of zone-to-zone transport already handled up front.
We also integrate a physical timing model that combines gate latencies with AOD motion delays to derive the overall execution time. This timing analysis feeds directly into our fidelity model, which accounts for gate errors, motion-induced infidelity, cross-talk on idle qubits inside the entanglement zone, and decoherence during idle periods—yielding a consistent estimate of circuit fidelity under the DT-based compilation flow.

\subsection{Customized Dynamic Compiler for QFT}

\noindent \textbf{Why QFT Requires Dynamic Channels and Why AOD Remains Essential.}
QFT-type circuits and Ising/QAOA-type circuits place fundamentally different demands on transport scheduling. In QFT, only a small and slowly shifting set of CZ pairs is active at each entangling stage. For a 10-qubit example, our ASAP scheduler generates 31 stages with 24 distinct but sparsely populated interaction patterns. Because only a few pairs are active simultaneously, provisioning a full set of static DT channels for all possible QFT interactions would be wasteful. Instead, the ``sliding-window’’ structure of QFT naturally motivates a dynamic approach in which a small number of DT channels are reshaped and reused across stages, covering the triangular CZ pattern without constructing a large, fully populated topology. In contrast, Ising/QAOA circuits repeatedly use the same dense interaction pattern and therefore benefit from a static-channel design.

Even with DT-based remote CZ, atomic transport remains necessary in zoned architectures. Qubits reside in SLM traps in the storage zone, while Rydberg gates can only be executed in the entanglement zone. AOD motion is thus required to (i) load qubits into the entanglement region, (ii) concentrate active qubits into a narrow cross-section suitable for DT routing, and (iii) repack ancilla and pre-use qubits along evolving DT paths. DT then propagates the Rydberg excitation along these paths to implement long-range CZ gates without relocating all data qubits. By treating AOD and DT as complementary primitives, our design avoids repeated global repacking before each entangling stage. Instead, AOD movement is limited to lightweight, local adjustments of the ancilla that define the DT channels, while the dominant transport burden is absorbed by fast DT propagation. This combination enables QFT’s stage-by-stage sliding pattern to be executed efficiently without constructing large static channels or incurring the heavy AOD overhead seen in prior zoned compilers.

\noindent\textbf{DT-Aware Compiler Flow}

\noindent \textit{DT eligibility and per-stage reuse.}
The compiler first determines if a two-qubit stage should use a DT channel. If CZ pairs persist across adjacent stages, DT planning is invoked to build a DT backbone; otherwise, the stage uses only local AOD-CZ. Once DT is enabled, the topology is incrementally updated. CZ pairs whose endpoints already lie near the channel, or can be placed there cheaply, are marked \emph{DT-eligible}; all others fall back to \emph{direct-AOD}. Qubits not involved in the stage are kept in the storage zone to limit AOD motion to the small set of active data qubits and channel ancilla.

\noindent \textit{Anchor-column selection on a weighted grid.}
As summarized in Algorithm~\ref{alg:dt-channel}, for a DT-enabled stage, channel construction is coupled with ancilla assignment using a weighted grid graph whose edge weights reflect AOD transport cost. From the current endpoints and the previous layout, the compiler selects a small set of anchor-column candidates consisting of the previous DT column band and the columns with the highest weighted endpoint density, and chooses the one minimizing the aggregate cost of connecting endpoint projections. A shortest-path search then constructs a DT backbone that passes through the region of concentrated endpoints while remaining close to the previous path.

\noindent \textit{Branch construction and ancilla assignment.}
Each endpoint must be connected to the DT backbone by a short, constraint-respecting branch. Branch paths avoid obstacles and expensive vertical moves. The union of backbone and branches defines the set of required channel sites. The compiler builds a compact cost matrix over available ancilla, including reused and reserved atoms, encoding AOD distance and reuse-preferred penalties, and solves a minimum-cost assignment via the Hungarian algorithm. The resulting move vectors are scheduled in parallel using MIS-based AOD routing.

\noindent \textit{Cross-stage reuse and fallback.}
DT planning incorporates endpoint rebalancing and cross-stage reuse: unchanged segments of the channel persist, and only conflicting sections are updated. If no feasible branch satisfies both motion and interaction constraints, the compiler uses a longer two-leg DT pattern or demotes the pair to \emph{direct-AOD}. This keeps the DT backbone stable across stages while accommodating QFT-style drifting interaction patterns.

\begin{algorithm}[t]
\footnotesize
\caption{DTChannelPlan for a DT-enabled 2Q stage $s$}
\label{alg:dt-channel}
\begin{algorithmic}[1]
\Require 2Qgate stage $s$;   CZ pair set $P$;   layout state $L$;
        previous DT channel $C_{\text{prev}}$;
        previous ancilla sets $A_{\text{ch}}^{\text{prev}}, A_{\text{rs}}^{\text{prev}}$
\Ensure updated DT channel $C$; DT pair set $P^{\text{dt}}$;
        AOD pair set $P^{\text{aod}}$

\State grid $\gets \textsc{BuildGrid}(L)$
%\State col$^\star \gets \textsc{PickCol}(P, C_{\text{prev}})$
%\State $C \gets \textsc{UpdateChan}(\text{grid}, \text{col}^\star, C_{\text{prev}})$
\State $C \gets \textsc{UpdateChan}(\text{grid}, C_{\text{prev}})$

\State $P^{\text{dt}}\gets\emptyset$, $P^{\text{aod}}\gets\emptyset$, sites$\gets\emptyset$
\For{each $(u,v)\in P$}
  \If{\textsc{NearChain}$(u,v,C)$ \textbf{ or } \textsc{CheapAlign}$(u,v,C)$}
    \State $P^{\text{dt}} \gets P^{\text{dt}} \cup \{(u,v)\}$
    \State $b_u \gets \textsc{FindBranch}(\text{grid}, \text{pos}(u), C)$
    \State $b_v \gets \textsc{FindBranch}(\text{grid}, \text{pos}(v), C)$
    \State sites $\gets$ sites $\cup b_u \cup b_v$
  \Else
    \State $P^{\text{aod}} \gets P^{\text{aod}} \cup \{(u,v)\}$
  \EndIf
\EndFor

\State anc $\gets A_{\text{ch}}^{\text{prev}} \cup A_{\text{rs}}^{\text{prev}}
            \cup \textsc{NewAnc}(L)$
\State cost $\gets \textsc{BuildCost}(\text{anc}, \text{sites})$
\State match $\gets \textsc{Hungarian}(\text{cost})$
\State \textsc{EmitMoves}(match)

\For{each $(u,v)\in P^{\text{dt}}$}
  \If{\textsc{TooExp}$(u,v)$ \textbf{ or } \textsc{Infeasible}$(u,v)$}
    \If{\textsc{TwoLegDT}$(u,v,C)$}
      \State \textsc{UseTwoLegDT}$(u,v)$
    \Else
      \State $P^{\text{dt}} \gets P^{\text{dt}} \setminus \{(u,v)\}$
      \State $P^{\text{aod}} \gets P^{\text{aod}} \cup \{(u,v)\}$
    \EndIf
  \EndIf
\EndFor
\end{algorithmic}
\end{algorithm}

\begin{figure*}
  \centering
  \includegraphics[width=0.94\linewidth]{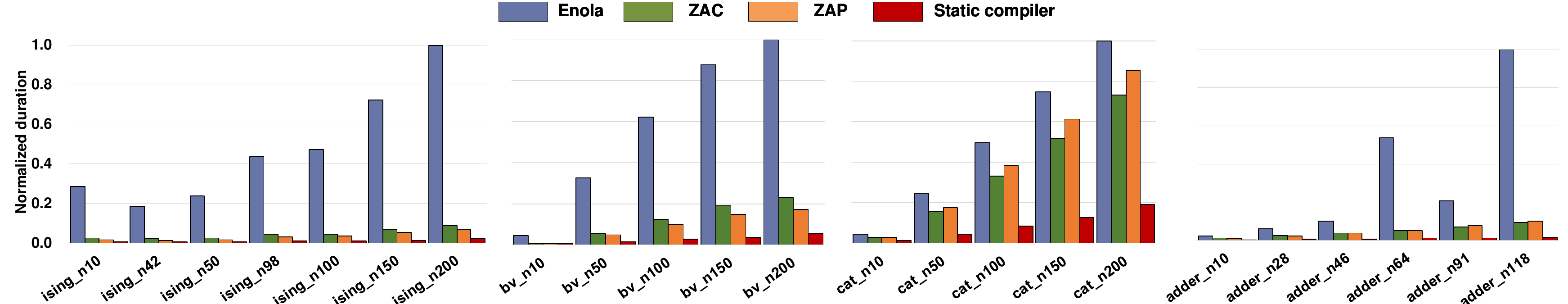}
  \caption{Duration comparison our proposed static compiler and SOTA compilers.}
  \label{fig:durationcomp}
  \vspace{-2.ex}
\end{figure*}

\begin{figure}
  \centering
  \includegraphics[width=0.68\linewidth]{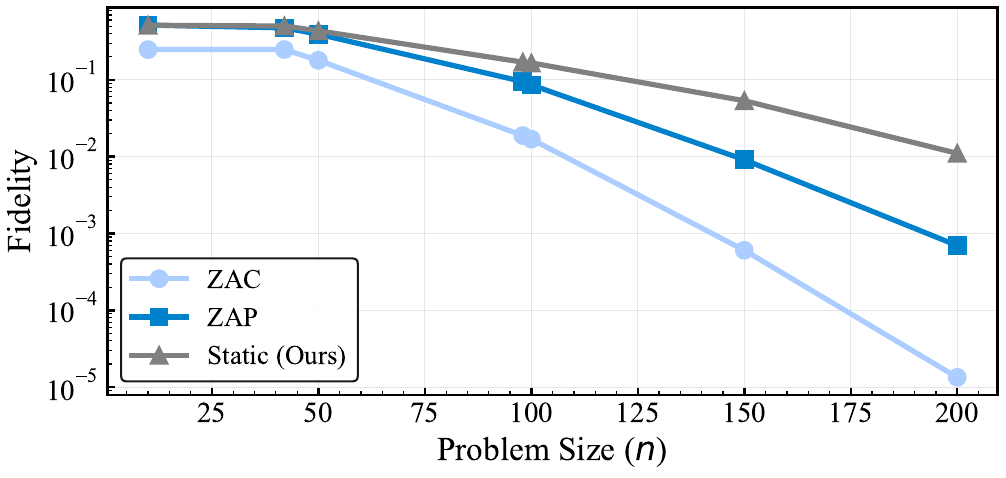}
  \vspace{-2mm}
  \caption{Fidelity comparison between our static compiler and SOTA compilers on different-sized Ising circuits.}
  \label{fig:fidelitycomp}
  \vspace{-2.ex}
\end{figure}

\begin{figure}
  \centering
  \includegraphics[width=0.93\linewidth]{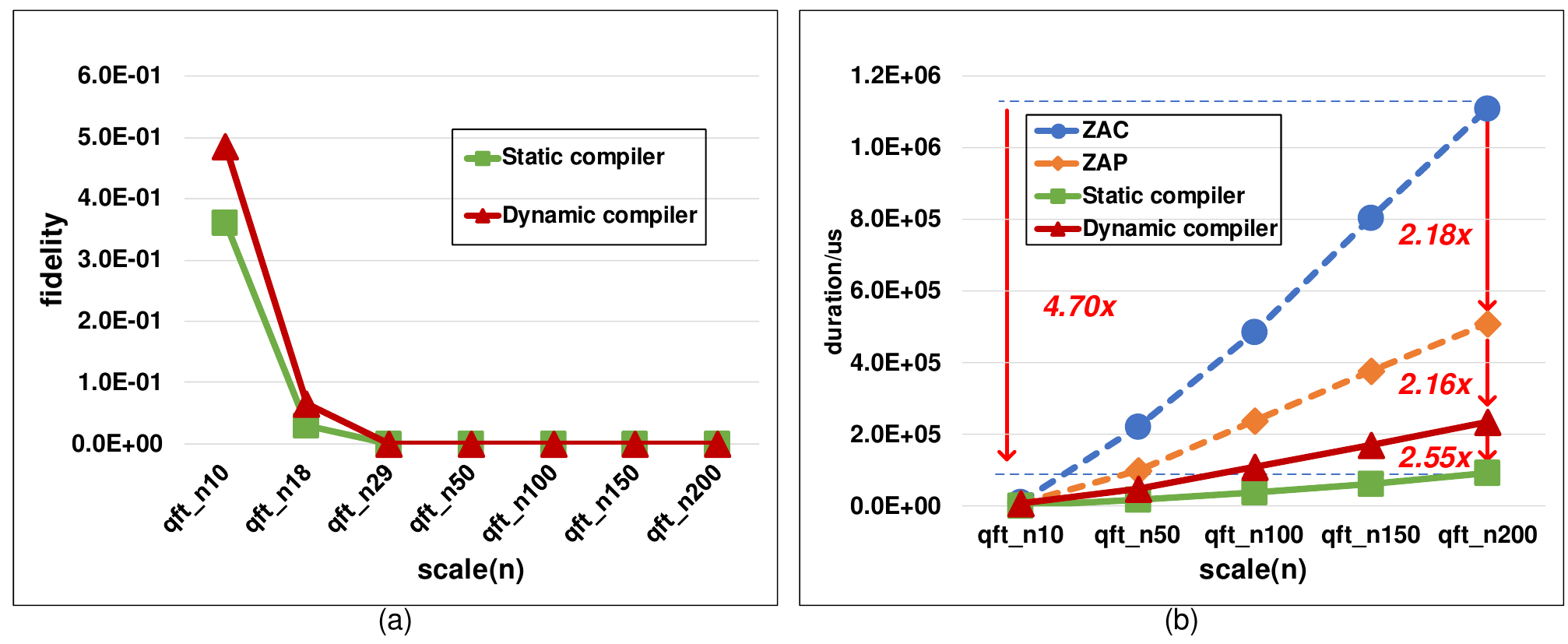}
  \caption{Our Proposed Static vs Dynamic Compiler: Performance comparison on different-sized QFT benchmarks.}
  %\Description{}
  \label{fig:staticdynamic}
  \vspace{-2.ex}
\end{figure}

\noindent\textbf{Reuse-Driven Ancilla and DT Channel Management.}
In the dynamic DT regime, a DT channel is built from two classes of physical resources: \emph{flying ancilla} and \emph{pre-use qubits}. Flying ancilla follow the FPQA model used in Q-Pilot~\cite{wang2024q}: any atom on an AOD site that is not storing logical information may be activated as a temporary ancilla, routed to mediate DT-based remote CZ gates, and then returned to the storage zone for reuse.

By contrast, \emph{pre-use qubits} are logical qubits that will eventually participate in the computation, but whose first logical gate appears only in a later portion of the circuit. Before their first logical use, these qubits are collectively loaded into the entanglement zone and treated as relay nodes along the DT channel, as illustrated by the ``pre-use'' segment in Fig.~\ref{fig:fpqa-flow}. Once the circuit reaches its first scheduled
logical gate, the compiler withdraws them from the channel and returns them to the normal data-qubit execution flow.

Within the flying-ancilla class, we distinguish two roles: \emph{channel ancilla} and \emph{reserve ancilla}. Channel ancilla occupy the active DT backbone in a two-qubit stage and form the physical sites along which the Rydberg excitation propagates. Reserve ancilla remain idle but are positioned so that they can be incorporated into the DT path at low AOD cost, providing a pool of ready relay sites when the channel must be extended, shortened, or rerouted around newly active qubits.

\section{Evaluation}

\noindent \textbf{Benchmarks.}
We evaluate our approach on five families of parameterized quantum circuits: Quantum Fourier Transform (\texttt{qft}), transverse-field Ising (\texttt{ising}), Bernstein--Vazirani (\texttt{bv}), cat-state preparation (\texttt{cat}), and ripple-carry adder (\texttt{adder}). 

\noindent \textbf{Baselines.}
We compare our compiler against three SOTA neutral-atom compilers: ZAC~\cite{lin2025reuse}, ZAP~\cite{huang2024zap}, and Enola~\cite{tan2025compilation}.
% \textbf{ZAC}, a reuse-aware compiler for zoned neutral-atom architectures; \textbf{ZAP}, a zoned architecture and parallelizable compiler; and \textbf{Enola}, a compiler for dynamically field-programmable atom arrays. 
To ensure a fair comparison, we adopt the same hardware parameter table and zoned architecture specification as ZAP.
% and report for all methods the two-qubit-gate depth, the total number of two-qubit gates, and the corresponding estimated execution time.

% \subsection{Main Results}
\noindent \textbf{Duration Comparision with SOTA Compilers:}
As shown in Fig.~\ref{fig:durationcomp}, our static compiler consistently reduces entangling-stage duration across all benchmark families compared with the best of Enola, ZAC, and ZAP. Across the \emph{Ising}, \emph{BV}, \emph{CAT}, and \emph{adder} suites, the minimum improvement observed is about $29\%$ (in the smallest BV instance), while the best cases reach $75$--$82\%$ reduction for large-scale circuits ($n\approx 150$--$200$). On average, the static compiler delivers a $72$--$74\%$ reduction in duration—corresponding to roughly $3.8$--$3.9\times$ speedup—uniformly outperforming all baselines. Importantly, the improvement grows with circuit size, indicating that the proposed static strategy scales favorably as circuits become deeper and wider.

\noindent \textbf{Fidelity Comparison with SOTA Compilers:}
Beyond reducing entangling duration, our static compiler also improves end-to-end fidelity on Ising benchmarks. As shown in Fig.~\ref{fig:durationcomp}, its total fidelity ranges from $0.513$ at $n{=}10$ to $0.011$ at $n{=}200$, whereas the best of ZAC and ZAP ranges from $0.515$ down to only $0.00070$. Averaged across all instances, the static compiler reaches $0.264$ fidelity versus $0.225$ for the best baseline, a $17.6\%$ improvement. At small sizes, our fidelity matches ZAP, but gains increase markedly with system size. For mid-range circuits ($n{=}42$--$50$), fidelity rises by $6$--$11\%$. For deeper circuits ($n{=}98$--$100$), the improvement reaches $1.8$--$1.9\times$. The largest gains occur for $n{=}150$--$200$, where our fidelities exceed the best baseline by factors of $5.9\times$ and $15.9\times$. Overall, the static DT strategy mitigates cumulative transport and decoherence errors, yielding substantially higher fidelity for large Ising workloads.

\noindent \textbf{Comparison of Static and Dynamic Compilers for QFT Tasks:}
Fig.~\ref{fig:staticdynamic} compares the static and dynamic DT compilers on QFT circuits. The dynamic compiler consistently achieves higher output fidelity, improving from $3.61\times10^{-1}$ to $4.85\times10^{-1}$ at small scales and from $1.67\times10^{-12}$ to $3.70\times10^{-7}$ in the most demanding cases. These results indicate that dynamic DT execution significantly enhances the robustness of long-range QFT workloads.

Fig.~\ref{fig:staticdynamic}(b) shows the DT-channel duration. Both compilers provide large reductions over Enola, ZAC, and ZAP. At $n{=}10$, the static compiler lowers the duration to $2.45\times10^{3}$, while the dynamic compiler maintains a still-low $7.19\times10^{3}$, well below prior methods. At $n{=}200$, the static and dynamic durations ($9.26\times10^{4}$ and $2.36\times10^{5}$, respectively) remain substantially smaller than those of ZAC and ZAP. Relative to the static baseline, the dynamic compiler accepts a moderate increase in DT duration in exchange for a substantial fidelity gain. 
The tradeoff is clear, the static compiler is optimized for minimal DT duration, whereas the dynamic compiler spends slightly more DT time to utilize DT channels more effectively, delivering higher-fidelity QFT execution while still achieving large duration reductions compared with existing compilers.\looseness=-1

\section{Conclusion And Outlook}
In this work, we introduce a hybrid remote-CZ compilation framework for zoned neutral-atom systems that tightly couples AOD-assisted configuration with DT execution. Our approach is the first to integrate DT channels into zoned architectures, enabling long-range CZ interactions with significantly reduced movement cost. We develop a static DT compiler for general workloads and a dynamic DT compiler optimized for QFT-style circuits, coordinated through lightweight AOD alignment inside the entanglement zone. 
Across diverse benchmarks, our method yields substantial reductions in entangling latency and, in many cases, improves output fidelity compared with existing zoned compilers. These results demonstrate that combining AOD and DT offers an effective and practical route for enhancing NISQ-era performance and scaling neutral-atom compilation to more complex quantum circuits.
% In this work, we introduce a hybrid remote-CZ compilation approach for zoned neutral-atom quantum computers that tightly couples AOD-assisted configuration with directional-transport execution. Our design is the first to explicitly integrate DT channels into zoned layouts, enabling long-range CZ interactions with substantially reduced movement overhead. We develop a static compiler for general workloads and a dynamic compiler tailored to QFT like tasks, coordinated through lightweight AOD alignment for zone-level repacking. Experiments across diverse benchmarks show clear reductions in entangling latency and even improvement in output fidelity compared with existing zoned compilers. This work highlights how combining AOD and DT can significantly improve NISQ-era performance and offers a practical path for scaling neutral-atom compilation to more complex circuits.
\enlargethispage{2\baselineskip}

\section{Acknowledgment}
The authors thank Qi‑Yu Liang for valuable discussions and insights.

\newpage

\bibliographystyle{ACM-Reference-Format}
\bibliography{ref}

\end{document}